\documentclass[twocolumn,prl,tightenlines,superscriptaddress]{revtex4}
\usepackage{epsfig}
\usepackage{graphics}
\usepackage{latexsym}
\usepackage{amsmath}
\usepackage{amssymb}
\usepackage{rotating}
\usepackage{color}

\begin{document}

\title{Solid quantization for non-point particles}
\author{P. Wang}
\affiliation{Institute of High Energy Physics, CAS, P. O. Box
918(4), Beijing 100049, China} \affiliation{Theoretical Physics
Center for Science Facilities, CAS, Beijing 100049, China}

\begin{abstract}
In quantum field theory, elemental particles are assumed to be point
particles. As a result, the loop integrals are divergent in many
cases. Regularization and renormalization are necessary in order to
get the physical finite results from the infinite, divergent loop
integrations. We propose new quantization conditions for non-point
particles. With this solid quantization, divergence could be treated
systematically. This method is useful for effective field theory
which is on hadron degrees of freedom. The elemental particles could
also be non-point ones. They can be studied in this approach as
well.

\end{abstract}


\maketitle
\smallskip

Quantum field theory is the fundamental theory for nuclear and
particle physics. The simplest way to quantize the field is to use
canonical quantization which is similar as in quantum mechanics. It
is equivalent to the path integral method. With the quantum field
theory, one can study the micro process with Feynman rules. When do
the high order calculation, the loop contribution will appear. These
integrals are often divergent, i.e., they become infinite when
momentum integration goes to infinity. This ultraviolet divergence
is short-distance phenomenon.

Many kinds of methods are introduced in quantum field theory to deal
with the divergence. One of the most popular method is dimensional
regularization \cite{Hooft}. It provides a systematic tool to obtain
finite physical results from the infinity. Another is Pauli-Villars
regularization which adds fictitious particles to the theory with
large masses to cancel out the infinity \cite{Pauli}.

Quantum field theory with dimensional regularization is very
standard and widely accepted. It is also applied in effective field
theory which is on hadron degrees of freedom
\cite{Gasser,Ecker,Meissner}. In hadron physics, there are a lot of
phenomenological models where divergence is often treated by adding
a cutoff or form factor to the integral ``by hand". The cutoff or
form factor can be related to the wave function which means hadrons
are not point particles \cite{Lu,Lyubovitskij}. It can also be
``derived" from the non-local interaction \cite{Ivanov,Hell}. In
other words, if particles are not point ones, there is no divergence
appear from the beginning. There exists quantum field theories for
point particles. Whether we can have some ``theories" for non-point
particles in which divergence can be avoided systematically? We will
show that from the new quantization conditions, one could get the
modified propagators for non-point particles. Divergence can be
treated systematically.

In fact, in the early 1950s, Yukawa has proposed the non-local
fields which described the non-point particles
\cite{Yukawa1,Yukawa2}. It was assumed that the non-local field was
a function of four space-time operators $x_\mu$ as well as of four
space-time displacement operators $p_\mu$. Besides the normal
equation, this field satisfied another one which was related to the
radius of elemental particle. However, this idea did not get widely
accepted because the aim to get rid of divergence was not easily
established \cite{Marnelius1,Moller}. In Refs.
\cite{Marnelius2,Imamura}, the authors claimed that there exists no
meaningful $S$ matrix with non-local interaction. While some authors
pointed out that the violation of unitary observed in space/time
noncommutative field theories was due to an improper definition of
quantum field theory on noncommutative spacetime (Quantum field
theory on the standard noncommutative spacetime is equivalent to a
non-local theory on a commutative spacetime.) \cite {Bahns}. As long
as a proper perturbative setup is employed, non-local field theories
may well be unitary in the sense that probabilities are always
conserved. A proof of unitary of $S$ matrix as well as causality in
a non-local quantum field theory has been shown in the paper of
Alebastrov and Efimov \cite{Alebastrov1,Alebastrov2}. At the same
time, non-local quantum electrodynamics was widely discussed
\cite{Efimov1,Efimov2,Efimov3,Efimov4,Phat,Terning}. In recent
years, a lot of work has been done on the non-local phenomenological
models as well as on the noncommutative field theory
\cite{Radzhabov,Dorokhov,Noguera,Faessler,Armoni,Balachandran}. For
practise, one can use the unitary operator $T \text
{exp}\{i\int_{-\infty} ^\infty d^4x {\cal L}_{int}(x)\}$, where
${\cal L}_{int}(x)$ is the non-local interaction, to do the
perturbative expansion order by order \cite{Ivanov,Faessler2}.

In this paper, we propose new quantization conditions for non-point
particles. Consistent with this solid quantization, the non-local
Lagrangian is straightforward. Different from the traditional
non-local case, here the free Lagrangian should be non-local as
well.

Let's start with the traditional canonical quantization for the
simplest scalar field. The traditional commutation relations are:
\begin{eqnarray}\nonumber
\left[\phi(\vec{x},t), \phi(\vec{y},t)\right] &=&
\left[\pi(\vec{x},t), \pi(\vec{y},t)\right] = 0,
\\
\left[\phi(\vec{x},t), \pi(\vec{y},t)\right] &=&
i\delta^{(3)}\left(\vec{x}-\vec{y}\right).
\end{eqnarray}
The $\delta$ function in the above equation means that a point
particle and anti-particle can only be created at the same position
point.

The field and its conjugate partner can be expanded in momentum
space, expressed as
\begin{equation}\label{expand}
\phi(\vec{x},t) = \int\widetilde{dp}
\left[a(\vec{p})e^{i\vec{p}\cdot\vec{x}-i\omega_pt} + a^\dag
(\vec{p})e^{-i\vec{p}\cdot\vec{x}+i\omega_pt}\right],
\end{equation}

\begin{equation}
\pi(\vec{x},t) = \int
\widetilde{dp}(-i)\omega_p\left[a(\vec{p})e^{i\vec{p}\cdot\vec{x}-i\omega_pt}
- a^\dag (\vec{p})e^{-i\vec{p}\cdot\vec{x}+i\omega_pt}\right],
\end{equation}
where
\begin{equation}
\widetilde{dp} = \frac{d^3p}{(2\pi)^3 2\omega_p}.
\end{equation}

It is straightforward to obtain the commutation relations between
creation and annihilation operators:
\begin{eqnarray} \nonumber
\left[a(\vec{p}), a(\vec{q})\right] &=& \left[a^\dag(\vec{p}),
a^\dag(\vec{q})\right] = 0, \\
\left[a(\vec{p}), a^\dag(\vec{q})\right] &=&
(2\pi)^32\omega_p\delta^{(3)} (\vec{p} - \vec{q}).
\end{eqnarray}
The creation operator creates a momentum state $|p\rangle =
a^\dag(\vec{p})|0\rangle $ which is normalized as
\begin{equation}
\int \widetilde{dp} |p\rangle \langle p| = 1.
\end{equation}

Because the particle is assumed to be point particle (behaves like
$\delta$ function in position space), when expanded in momentum
space, it has the same possibility for different momentum. However,
the real particle could be like a wavepacket. It is partially
localized in both position and momentum space. The possibility of
the particle with high momentum is small. With high-momentum
suppression, the divergence in the loop integral may not appear.

Therefore, we propose new quantization conditions (solid
quantization):
\begin{eqnarray}\nonumber\label{sq}
\left[\phi(\vec{x},t), \phi(\vec{y},t)\right] &=&
\left[\pi(\vec{x},t), \pi(\vec{y},t)\right] = 0,
\\
\left[\phi(\vec{x},t), \pi(\vec{y},t)\right] &=&
i\Phi\left(\vec{x}-\vec{y}\right).
\end{eqnarray}
The function $\Phi\left(\vec{x}-\vec{y}\right)$ describes the
correlation between fields at $\vec{x}$ and $\vec{y}$. Due to the
fact that particle is not a dimensionless point particle, but a
solid one, particles at different positions could be partially
superimposed which means there exists some possibility that particle
and antiparticle are created in different positions.

One can also expand the field as Eq.~(\ref{expand}) (In this case,
we use capital letter $A$ instead of $a$.)
\begin{equation}
\phi(\vec{x},t) = \int\widetilde{dp}
\left[A(\vec{p})e^{i\vec{p}\cdot\vec{x}-i\omega_pt} + A^\dag
(\vec{p})e^{-i\vec{p}\cdot\vec{x}+i\omega_pt}\right].
\end{equation}
As a result, the creation and annihilation operators satisfy the
following relations
\begin{eqnarray}
\left[A(\vec{p}), A(\vec{q})\right] &=& \left[A^\dag(\vec{p}),
A^\dag(\vec{q})\right] = 0, \nonumber \\
\left[A(\vec{p}), A^\dag(\vec{q})\right] &=&
(2\pi)^32\omega_p\delta^{(3)} (\vec{p} - \vec{q})\Psi (\vec{p}).
\end{eqnarray}
$\Phi\left(\vec{x}\right)$ and $\Psi\left(\vec{p}\right)$ obey the
following relations
\begin{equation}
\Phi\left(\vec{x}\right)=\int
\frac{d^3p}{(2\pi)^3}\frac{\Psi(\vec{p})}{2}(e^{i\vec{p}\cdot\vec{x}}+e^{-i\vec{p}\cdot\vec{x}}),
\end{equation}
\begin{equation}
\Psi\left(\vec{p}\right)=\int
d^3x\frac{\Phi(\vec{x})}{2}(e^{i\vec{p}\cdot\vec{x}}+e^{-i\vec{p}\cdot\vec{x}}).
\end{equation}

The above two equations generate two normalization formulas
\begin{equation}
\Phi(0)=\int\frac{d^3p}{(2\pi)^3}\Psi(\vec{p}),
\end{equation}
\begin{equation}
\Psi(0)=\int d^3x\Phi(\vec{x})=1.
\end{equation}
Compared with the traditional commutation relation where
$\Phi\left(\vec{x}\right)=\delta^{(3)}(\vec{x})$,
$\Phi\left(\vec{x}\right)$ is normalized to be 1, while
$\Psi(\vec{p})$ is normalized to be $\Phi(0)$.

With the new quantization, the field can be written in terms of
traditional creation and annihilation operators as
\begin{equation}
\phi(\vec{x},t) = \int \widetilde
{dp}\sqrt{\Psi(\vec{p})}\left[a(\vec{p})e^{i\vec{p}\cdot\vec{x}-i\omega_pt}
+ a^\dag (\vec{p})e^{-i\vec{p}\cdot\vec{x}+i\omega_pt}\right].
\end{equation}

It is easy to get the Feynman propagator of the scalar field in the
solid quantization. The propagator is defined as
\begin{eqnarray}\nonumber
&& \Delta_F (x'-x) = \langle 0|T\phi(x')\phi(x)|0\rangle \\
&& = \int\widetilde{dk}\left[\theta(t'-t)e^{ik\cdot(x'-x)}
+\theta(t-t')e^{-ik\cdot(x'-x)}\right].~~~~~~
\end{eqnarray}
The integral expression of the step function is
\begin{equation}
\theta(t)= \text {lim}_{\epsilon\rightarrow 0^+} \int
\frac{d\tau}{2\pi i} \frac{e^{i\tau t}}{\tau-i\epsilon}.
\end{equation}
With the help of the above equation, the Feynman propagator can be
obtained as
\begin{equation}
\Delta_F (x'-x) = \int
\frac{d^4k}{(2\pi)^4}\frac{i\Psi(\vec{k})e^{-ik\cdot(x'-x)}}{k^2 -
m^2 + i\epsilon}.
\end{equation}

For the other fields, the quantization condition is similar. For
example, for spin 1/2 fermion, the nonzero anti-commutation
relationship is
\begin{equation}
\left\{\psi_\alpha(\vec{x},t), \bar{\psi}_\beta(\vec{y},t)\right\} =
\gamma^0_{\alpha\beta} \Phi(\vec{x}-\vec{y}).
\end{equation}
Correspondingly, the field should be written as
\begin{eqnarray}\nonumber
\psi(\vec{x},t) &=&
\sum_{s=\pm}\int\widetilde{dp}\sqrt{\Psi(\vec{p})}
\left[b_s(\vec{p})u_s(\vec{p})e^{i\vec{p}\cdot\vec{x}-i\omega_pt}
\right . \\
&& \left . + d_s^\dag
(\vec{p})v_s(\vec{p})e^{-i\vec{p}\cdot\vec{x}+i\omega_pt}\right],
\end{eqnarray}
where $b$ and $d^\dag$ are normal annihilation and creation
operators. $u_s(\vec{p})$ and $v_s(\vec{p})$ are Dirac spinors. The
propagator of the spin 1/2 field can be obtained as
\begin{equation}
S_F (x'-x) = \int
\frac{d^4k}{(2\pi)^4}\frac{i\Psi(\vec{k})(k\cdot\gamma+m)e^{-ik\cdot(x'-x)}}{k^2
- m^2 + i\epsilon}.
\end{equation}

Vector field, say photon field can also be expanded as
\begin{eqnarray}\nonumber
A^\mu(\vec{x},t) &=&
\sum_{\lambda=\pm}\int\widetilde{dp}\sqrt{\Psi(\vec{p})}
\left[a_\lambda(\vec{p})\epsilon^\mu(\vec{p},\lambda)e^{i\vec{p}\cdot\vec{x}-i\omega_pt}
\right . \\
&& \left . + a^\dag_\lambda
(\vec{p})\epsilon^{*\mu}(\vec{p},\lambda)e^{-i\vec{p}\cdot\vec{x}+i\omega_pt}\right],
\end{eqnarray}
where $\epsilon^\mu(\vec{p},\lambda)$ is the polarization vector.
The photon propagator can be written as
\begin{equation}
D_F^{\mu\nu}(x'-x) = \int
\frac{d^4k}{(2\pi)^4}\frac{-i\Psi(\vec{k})g^{\mu\nu}e^{-ik\cdot(x'-x)}}{k^2
- m^2 + i\epsilon}.
\end{equation}

We should mention that, in principle, the function $\Psi(\vec{p})$
or $\Phi(\vec{x}-\vec{y})$ is particle dependent. It describes the
particle's property in addition to the mass and width. Therefore,
with the new quantization conditions, the Feynman rules should be
changed correspondingly. The new propagator of the field is
multiplied by a factor $\Psi(\vec{k})$ and the external field is
multiplied by a factor $\sqrt{\Psi(\vec{k})}$.

A question may arise here that how to connect the new propagator
with the path integral formulation. The path integral for the free
point-like field is defined as
\begin{equation}
Z_0(J) = \int {\cal D} \phi e^{i\int d^4x [{\cal L}_0 + J\phi]},
\end{equation}
where
\begin{equation}
{\cal L}_0 = -\frac12 \partial^\mu\phi\partial_\mu\phi-\frac12
m^2\phi^2
\end{equation}
is the Lagrangian density and $J$ is the external current. For a
solid particle, the free Lagrangian density is different. The
density can be written as
\begin{equation}\label{l0}
{\cal L}_0 = \phi\frac{(\partial^\mu\partial_\mu-
m^2)}{2\Psi(i\vec{\partial})}\phi.
\end{equation}
With the above Lagrangian density, the propagator of scalar field
obtained in the path integral formulation is the same as that in
solid canonical quantization. For the fermion and vector fields, the
situation is the same.

The factor $\Phi(\vec{x}-\vec{y})$ is the correlation of two
particles at $\vec{x}$ and $\vec{y}$. If we choose
$\Phi(\vec{x}-\vec{y}) = \delta^{(3)}(\vec{x}-\vec{y})$,
$\Psi(\vec{p})$ will equal 1. All of the above propagators will be
changed back to the conventional ones. As we explained previously,
the particle could be a solid particle with three space dimensions.
The particle and antiparticle can be created at small distance.
Therefore, the function of $\Phi(\vec{x}-\vec{y})$ can be a function
which decreases with the increasing distance $|\vec{x}-\vec{y}|$.
The smaller the particle, the closer the function to $\delta$
function.

The above solid quantization provides the ``kinematics" for quantum
field theory. Now let's look at the ``dynamics" for non-point
particles. Gauge invariance is a fundamental method to get the
strong or electro-weak interactions. Since particles are not point
ones, in general, the interaction among them is non-local. Similar
as the non-local quark-meson interaction \cite{Ivanov,Faessler2},
the gauge invariant interaction between fermion and gauge field, say
photon field, can be written as
\begin{eqnarray}\label{lt}
{\cal L}(x) &=& \int d^3a \bar{\psi}(t,\vec{x}+ \frac
{\vec{a}}{2})e^{iI(\vec{x}+\vec{a}/2,\vec{x})}\gamma^\mu iD_\mu
 \nonumber \\
&& e^{-iI(\vec{x}-\vec{a}/2,\vec{x})}\psi(t,\vec{x}- \frac
{\vec{a}}{2}) F(\vec{a}),
\end{eqnarray}
where $D_\mu = \partial_\mu - ig\int d^3 b
A_\mu(t,\vec{x}+\vec{b})G(\vec{a},\vec{b})$ and $I(y,x)= g\int_x^y
dz_\mu \int d^3 b A^\mu(z_0,\vec{z}+\vec{b})G(\vec{a},\vec{b})$. $g$
is the coupling constant. The function $F(\vec{a})$ and
$G(\vec{a},\vec{b})$ are related to the size of fermion and gauge
fields. The non-local coupling depends on the distance between the
two fermion fields and the distance between gauge field and the
center of two fermion fields. The coupling $G(\vec{a},\vec{b})$ can
be factorized as $\frac{\Phi(\vec{a})\Phi_g(\vec{b})}{F(\vec{a})}$,
where $\Phi(\vec{a})$ is the correlation function between two
fermions at distance $\vec{a}$ defined in Eq.~(\ref{sq}).
$\Phi_g(\vec{b})$ the correlation function for gauge fields. The
particular choice of $G(\vec{a},\vec{b})$ is to get the interacting
term $\int d^3a\int d^3b \bar{\psi}(t,\vec{x}+ \frac
{\vec{a}}{2})\gamma^\mu \psi(t,\vec{x}- \frac {\vec{a}}{2})
A_\mu(t,\vec{x}+\vec{b})\Phi(\vec{a})\Phi_g(\vec{b})$ which provides
the possibility $\Phi(\vec{a})\Phi_g(\vec{b})$ for the non-local
interaction.

The above Lagrangian is invariant under the following gauge
transformation:
\begin{eqnarray}\nonumber
\psi(t,\vec{x}-\vec{a}/2)\rightarrow
e^{ig_{\text{eff}}\theta(t,\vec{x}-\vec{a}/2)}\psi(t,\vec{x}-\vec{a}/2),~~~~~~~\\
\nonumber \bar{\psi}(t,\vec{x}+\vec{a}/2)\rightarrow
\bar{\psi}(t,\vec{x}+\vec{a}/2)e^{-ig_{\text{eff}}\theta(t,\vec{x}+\vec{a}/2)},~~~~~ \\
A_\mu(t,\vec{x}+\vec{b})\rightarrow A_\mu
(t,\vec{x}+\vec{b})+\partial_\mu \theta'(t,\vec{x}+\vec{b}),~~~~~
\end{eqnarray}
where $g_{\text{eff}}$ is defined as
$\frac{\Phi(\vec{a})\Phi_g(\vec{b})}{F(\vec{a})}$ which can be
understood as the effective charge of a non-local electromagnetic
current with a distance $\vec{a}$ between a fermion and an
anti-fermion. The appearance of  $F(\vec{a})$ in the denominator of
$g_{\text{eff}}$ is because of the non-point property of the
fermions. The functions $\theta$ and $\theta'$ have the following
relation
\begin{equation}
\theta(t,\vec{x})=\int d^3 b \theta'(t,\vec{x}+\vec{b}).
\end{equation}
The strong and weak interaction can be easily obtained
in the same way with $SU(3)$ and $SU(2)$ generators.

The free Lagrangian density without gauge field is
\begin{equation}
{\cal L}_0(x)= \int d^3a \bar{\psi}(t,\vec{x}+ \frac
{\vec{a}}{2})\gamma^\mu i\partial_\mu \psi(t,\vec{x}- \frac
{\vec{a}}{2})F(\vec{a}).
\end{equation}
This non-local free Lagrangian is different from that in the
traditional non-local models where the free part of the Lagrangian
is local and the interaction part is a non-local coupling of point
particles
\cite{Ivanov,Radzhabov,Dorokhov,Noguera,Faessler,Faessler2}. Our
treatment is more consistent. Due to the non-point assumption, the
non-local Lagrangian is straightforward and it is also necessary
because of the solid quantization. After moving the position to the
same point by the translation operator, it is straightforward to
rewrite the above Lagrangian as
\begin{equation}\label{l0f}
{\cal L}_0(x)= \bar{\psi}(x)\gamma^\mu i\partial_\mu
\tilde{F}(i\vec{\partial})\psi(x),
\end{equation}
where $\tilde{F}(i\vec{\partial})$ is the Fourier transformation of
of $F(\vec{a})$, i.e.,
\begin{equation}
\tilde{F}(i\vec{\partial}) = \int d^3a e^{i\vec{a}\cdot
i\vec{\partial}} F(\vec{a}).
\end{equation}
Comparing Eqs.~(\ref{l0}) and (\ref{l0f}), we can get the
relationship $\tilde{F}(i\vec{\partial}) = 1/\Psi(i\vec{\partial})$.
One can see that the solid quantization is consistent with the path
integral approach with non-local Lagrangian density.

The interaction term is written as
\begin{eqnarray}
{\cal L}_{int}(x)&=& g \int d^3a \int d^3b
\bar{\psi}(t,\vec{x}+\frac{\vec{a}}{2})e^{iI(\vec{x}+\vec{a}/2,\vec{x})}\gamma^\mu
\nonumber \\
&& A_\mu(t,\vec{x}+\vec{b})
e^{-iI(\vec{x}-\vec{a}/2,\vec{x})}\psi(t,\vec{x}-\frac
{\vec{a}}{2})\Phi(\vec{a})\Phi_g(\vec{b})
\nonumber \\
&+& \int d^3a
\bar{\psi}(t,\vec{x}+\frac{\vec{a}}{2})(e^{iI(\vec{x}+\vec{a}/2,\vec{x})}-1)\gamma^\mu
i\partial_\mu \nonumber \\
&& e^{-iI(\vec{x}-\vec{a}/2,\vec{x})}\psi(t,\vec{x}-\frac
{\vec{a}}{2})F(\vec{a}) \nonumber \\
&+& \int d^3a \bar{\psi}(t,\vec{x}+\frac{\vec{a}}{2})\gamma^\mu
i\partial_\mu (e^{-iI(\vec{x}-\vec{a}/2,\vec{x})}-1) \nonumber \\
&& \psi(t,\vec{x}-\frac {\vec{a}}{2})F(\vec{a}).
\end{eqnarray}
The field can be expanded in power of $\vec{a}$ and $\vec{b}$ as
\begin{eqnarray}
\psi(t,\vec{x}+\vec{a}) =
\psi(t,\vec{x})+\vec{\partial}\psi(t,\vec{x})\cdot \vec{a} + {\cal
O}(\vec{a}^2), \nonumber \\
A_\mu(t,\vec{x}+\vec{b}) =
A_\mu(t,\vec{x})+\vec{\partial}A_\mu(t,\vec{x})\cdot \vec{b} + {\cal
O}(\vec{b}^2).
\end{eqnarray}
The interaction can be expressed as
\begin{equation}\label{int}
{\cal L}_{int}(x)=g\bar{\psi}(x)\gamma^\mu A_\mu(x)\psi(x) + {\cal
O}(\bar{\vec{a}},\bar{\vec{b}}),
\end{equation}
where $\bar{\vec{a}}$ and $\bar{\vec{b}}$ reflect the size of the
particles defined as
\begin{eqnarray} \nonumber
&&\overline{\vec{a}} = \int d^3 a ~\vec{a}~\Phi(\vec{a}),\\
&&\overline{\vec{b}} = \int d^3 b ~\vec{b}~\Phi_g(\vec{b}).
\end{eqnarray}

For the ``free" Lagrangian, we should not expand it in terms of
$\vec{a}$. The ``free" Lagrangian provides propagators for solid
particles. The further volume effect of solid particles can be added
order by order. If the particle's size is small enough, we can
neglect high order terms in Eq.~(\ref{int}). The lowest order
interaction term is the same as that in the local case.

The solid quantization is valid for elemental particles as well as
for hadrons if elemental particles are not point ones either. With
the new propagator, the loop integration is convergent. For example,
let's look at the following integration which appears in the photon
self-energy at one-loop level:
\begin{equation}
I = \int
\frac{d^4k}{(2\pi)^4}\frac{\Psi(\vec{k})\Psi(\vec{k}+\vec{p})}{\left[k^2-m^2\right]\left[(p+k)^2-m^2\right]},
\end{equation}
where $p$ is the external momentum of photon. $k$ and $k+p$ are the
internal momentum of two electron or quark propagators. After
integration of $k_0$, the above equation can be written as
\begin{eqnarray}\nonumber
I &=& \int
\frac{d^3k}{2(2\pi)^3}\left\{\frac{-i\Psi(\vec{k})\Psi(\vec{k}+\vec{p})}{\omega(\vec{k})
\left[(\omega(\vec{k})+\omega(\vec{p}))^2
-\omega^2(\vec{k}+\vec{p})\right]} - \right . \\
&& \left .
\frac{i\Psi(\vec{k})\Psi(\vec{k}+\vec{p})}{\omega(\vec{k}+\vec{p})
\left[(\omega(\vec{k}+\vec{p})-\omega(\vec{p}))^2-\omega^2(\vec{k})\right]}\right\},
\end{eqnarray}
where $\omega(\vec{q})=\sqrt{\vec{q}^2+m^2}$. Without the factor
$\Psi(\vec{k})$ and $\Psi(\vec{k}+\vec{p})$, the above integration
is log-divergent. Since the particle is a solid one with three
dimensions, its wave-function is suppressed at high momentum. If we
choose $\Psi(\vec{k})$ to be a dipole or Gauss function, the
integration is convergent.

Without renormalization, the running coupling constant can also be
understood. With the new quantization conditions, even at tree
level, the coupling constant will be associated with a momentum
dependent factor. For example, for the fermion-boson coupling, if
the initial and final momentum of fermions are $-\vec{q}/2$ and
$\vec{q}/2$, the momentum of the boson is $\vec{q}$, and the
momentum dependent factor of the coupling constant at tree level is
$\sqrt{\Psi_f(-\vec{q}/2)\Psi_f(\vec{q}/2)\Psi_g(\vec{q})}$. The
labels $f$ and $g$ are for fermion and gauge boson, respectively.
The asymptotic free is a general property not only for strong
interaction. It is because the particle is not a point one. The
momentum is partially localized which favors at low value.

Investigating quantum electrodynamic process is a good and clean way
to test this quantization for elemental particles. Let's study the
electron-photon Compton scattering for an example:
\begin{equation}
e^-(p,s) + \gamma(k,\epsilon) \rightarrow e^-(p',s') +
\gamma(k',\epsilon'),
\end{equation}
where $p$ and $p'$, $k$ and $k'$ are the initial and final momentum
of electron and photon, respectively. $s$ and $\epsilon$ are their
spin and polarization. The scattering amplitude can be obtained as
\begin{eqnarray}\nonumber
&& {\cal M} =
-e^2\sqrt{\Psi_e(\vec{p})\Psi_e(\vec{p'})\Psi_\gamma(\vec{k})\Psi_\gamma(\vec{k'})}
\bar{u}_{s'}(\vec{p}')\left[\Psi_e(\vec{p}+\vec{k}) \right . \\
\nonumber && \left . \not\epsilon'^*\frac{1}{\not p+\not k -
m}\not\epsilon +\Psi_e(\vec{p}-\vec{k'})\not\epsilon\frac{1}{\not
p-\not k' - m}\not\epsilon'^*\right]u_s(\vec{p}). \\
\end{eqnarray}
In the Lab frame where the initial electron is at rest, after
summing over the initial and final electron spins, averaged square
of amplitude is simplified as
\begin{equation}
\bar{{\cal M}^2} =
\frac{e^4}{8m^2}\left[\frac{A}{\omega^2}+\frac{B}{\omega'^2}
+\frac{C}{\omega\omega'}\right],
\end{equation}
where $\omega$ and $\omega'$ are energy of initial and final photon.
$A$, $B$ and $C$ is expressed as
\begin{equation}
A=8m\omega\left[2(k\cdot\epsilon')^2+m\omega'\right]F_1(\omega,\omega'),
\end{equation}
\begin{equation}
B=-8m\omega'\left[2(k'\cdot\epsilon)^2-m\omega\right]F_2(\omega,\omega'),
\end{equation}
\begin{eqnarray}\nonumber
C&=&\left[16m^2\omega\omega'[2(\epsilon\cdot\epsilon')^2-1]-16m\omega'(k\cdot\epsilon')^2
\right . \\
&& \left . +16m\omega(k'\cdot\epsilon)^2\right]F_3(\omega,\omega'),
\end{eqnarray}
where
\begin{eqnarray}\nonumber
F_1(\omega,\omega')&=&
\Psi_e(\omega^2+\omega'^2-2\omega\omega'cos\theta)\Psi_\gamma(\omega^2)
\\ \nonumber
&& \Psi_\gamma(\omega'^2)\Psi_e^2(\omega^2), \\ \nonumber
F_2(\omega,\omega')&=&
\Psi_e(\omega^2+\omega'^2-2\omega\omega'cos\theta)\Psi_\gamma(\omega^2)
\\ \nonumber
&& \Psi_\gamma(\omega'^2)\Psi_e^2(\omega'^2), \\ \nonumber
F_3(\omega,\omega')&=&
\Psi_e(\omega^2+\omega'^2-2\omega\omega'cos\theta)\Psi_\gamma(\omega^2)
\\
&& \Psi_\gamma(\omega'^2)\Psi_e(\omega^2)\Psi_e(\omega'^2)
\end{eqnarray}
are the additional functions associated with the new quantization
and $\theta$ is the angle between initial and final momentum of
photon. With $\alpha=e^2/4\pi$, the differential cross section can
then be written as
\begin{equation}
\frac{d\sigma}{d\Omega}=\frac{\alpha^2
}{32m^4}\left(\frac{\omega'}{\omega}\right)^2
\left[\frac{A}{\omega^2}+\frac{B}{\omega'^2}
+\frac{C}{\omega\omega'}\right].
\end{equation}
In the point particle approximation, the function of $\Psi(\vec{p})$
equals 1 and the above cross section is changed back to the
traditional one
\begin{equation}
\frac{d\sigma}{d\Omega}=\frac{\alpha^2
}{4m^2}\left(\frac{\omega'}{\omega}\right)^2
\left[\frac{\omega'}{\omega}+\frac{\omega}{\omega'}
+4(\epsilon\cdot\epsilon')^2-2\right].
\end{equation}

To test the solid quantization, it is interesting to measure cross
section of high energy electron-photon Compton scattering because
the function $\Psi_e$ and $\Psi_\gamma$ will have significant
decrease at high momentum (energy). The smaller the particle, the
larger the energy at which cross section has a clear difference from
the traditional one.

Due to the inclusion of the size of the particle, the above fields
($\phi(x)$ or $\psi(x)$) as well as the propagators are not Lorentz
covariant quantities. Our start point is that at each time $t$, each
particle has a distribution on space. It is obviously
non-relativistic though this physical picture is very clear and
similar to many phenomenological models. It is also easy for us to
apply this approach in numerical calculation. For example, in the
effective field theory, finite regularization in which a $\vec{k}^2$
dependent regulator $u(\vec{k}^2)$ was introduced ``by hand" was
used to get rid of the divergence \cite{Leinweber,Wang1}. We can use
the above solid propagators for hadrons to investigate the meson
loop contribution. Compared with finite range regularization, this
approach automatically gives the ``regulator" for each diagram. The
obtained ``regulator" is diagram dependent.

Now we give the relativistic version of the solid quantization.
Different from the non-relativistic case, the field has a
distribution on four dimensional space-time. For a scalar field
$\phi(x)$, it can be written as
\begin{equation}
\phi(x)=\int \frac{d^4p}{(2\pi)^4} H(p^2)\left[\alpha_p e^{-ip\cdot
x}+ \alpha_p^\dag e^{ip\cdot x}\right].
\end{equation}
The operators $\alpha_p$ and $\alpha_p^\dag$ have the following
commutation relations:
\begin{eqnarray} \nonumber
\left[\alpha_p, \alpha_q\right] &=& \left[\alpha_p^\dag,\alpha_q^\dag\right] = 0, \\
\left[\alpha_p,\alpha_q^\dag\right] &=& (2\pi)^4\delta^{(4)} (p -
q).
\end{eqnarray}
The commutation relations of scalar field and its conjugate are
\begin{eqnarray}\nonumber
\left[\phi(\vec{x},t),\pi(\vec{y},t)\right]&=&\int\frac{d^4p}{(2\pi)^4}H^2(p^2)ip_0(e^{i\vec{p}\cdot\vec{x}}+e^{-i\vec{p}\cdot\vec{x}})\\
\nonumber &=& \int\frac{d^3p}{(2\pi)^3}\frac{i\Psi(\vec{p})}{2}(e^{i\vec{p}\cdot\vec{x}}+e^{-i\vec{p}\cdot\vec{x}})\\
&\equiv& i\Phi(\vec{x}-\vec{y}),
\end{eqnarray}
where
\begin{equation}
\Psi(\vec{p})=\int\frac{dp_0}{\pi}H^2(p^2)p_0.
\end{equation}
For point particle with mass $m$, $\Psi(\vec{p})=1$ and
$H^2(p^2)=2\pi\delta(p^2-m^2)$. We should mention that $H(p^2)$ is
proportional to $\delta^{1/2}(p^2-m^2)$ instead of
$\delta(p^2-m^2)$. This is because the field is expanded in terms of
$\alpha_p$ and $\alpha^\dag_p$ instead of $a_p$ and $a_p^\dag$.

For simplicity, we rewrite the scalar field as
\begin{eqnarray}\nonumber
\phi(x)&=&\int\frac{d^4p}{(2\pi)^4}dM^2H(M^2)\delta(p^2-M^2)
\\ \nonumber &&
\left[\alpha_p e^{-ip\cdot x}+ \alpha_p^\dag e^{ip\cdot x}\right] \\
\nonumber &=&\int\frac{d^3p}{(2\pi)^42\omega_M}dM^2H(M^2)
\\ &&
\left[\alpha_{\vec{p},\omega_M} e^{i\vec{p}\cdot\vec{x}-i\omega_Mt}+
\alpha_{\vec{p},\omega_M}^\dag
e^{-i\vec{p}\cdot\vec{x}+i\omega_Mt}\right],
\end{eqnarray}
where $\omega_M=\sqrt{\vec{p}^2+M^2}$.

We can get the propagator of scalar field as
\begin{eqnarray}\ \nonumber
\Delta_F(x'-x)=\int\frac{d^3k}{4(2\pi)^4\omega_M\omega_{M'}}dM^2
dM'^2 H(M^2)H(M'^2) &&
\\
\delta(\omega_{M'}-\omega_M)
\left[\theta(t'-t)e^{ik\cdot(x'-x)}+\theta(t-t')e^{ik\cdot(x-x')}\right],
~~~ &&
\end{eqnarray}
where $\delta(\omega_{M'}-\omega_M)=2\omega_M\delta(M'^2-M^2)$. With
the definition of $\theta$ function, the propagator can be written
as
\begin{equation}
\Delta_F(x'-x)=\int\frac{d^4k}{(2\pi)^4}\frac{dM^2}{2\pi}\frac{iH^2(M^2)}{k^2-M^2+i\epsilon}e^{-ik\cdot(x'-x)}.
\end{equation}
Again, if $H^2(M^2)=2\pi\delta(M^2-m^2)$, the propagator is the same
as that for point particle with mass $m$. If $H^2(M^2)$ is chosen to
be $2\pi[\delta(M^2-m^2)-\delta(M^2-\Lambda^2)]$, one can get
Pauli-Villars regularization.

In the relativistic case, the Lagrangian density can be written in
the same way as Eq.~(\ref{lt}) except the integral is on four
dimensional space-time because both time and space are non-local,
i.e.
\begin{eqnarray}\label{ltr}
{\cal L}(x) &=& \int d^4a \bar{\psi}(x+\frac
{a}{2})e^{iI(x+a/2,x)}\gamma^\mu iD_\mu
 \nonumber \\
&& e^{-iI(x-a/2,x)}\psi(x-\frac{a}{2}) F(a),
\end{eqnarray}
where $D_\mu = \partial_\mu - ig\int d^4 b
A_\mu(x+b)\frac{\Phi(a)\Phi_g(b)}{F(a)}$ and $I(y,x)= g\int_x^y
dz_\mu \int d^4 b A^\mu(z+b)\frac{\Phi(a)\Phi_g(b)}{F(a)}$. In the
relativistic case, the function $\Phi(a)$ or $\Phi_g(b)$ is defined
to be the Fourier transformation of $1/\tilde {F}(p^2)$, i.e.
\begin{equation}
\Phi(a)=\int \frac{d^4 p}{(2\pi)^4}  \frac{1}{\tilde
{F}(p^2)}e^{ip\cdot a},
\end{equation}
where $\tilde {F}(p^2)$ is the Fourier transformation of function
$F(a)$. Similar as in the non-relativistic case, by comparing with
the propagators obtained in solid quantization and path integral
methods, one can get the relationship between $H(p^2)$ and $\tilde
{F}(p^2)$:
\begin{equation}
\int \frac{dM^2}{2\pi}\frac{H^2(M^2)}{p^2-M^2} = \frac{1}{\tilde
{F}(p^2)(p^2-m^2)}.
\end{equation}
Again, one can see that, to be consistent with the solid
quantization, the Lagrangian is non-local in both free and
interaction parts.

Due to the non-local property, the conversation laws are modified
accordingly \cite{Garczynski}. The currents or charges are not
conserved at any space-time point. But the integral of them are
conserved. For example, in the non-local case, there exists no
unitary time evolution operator $U(t_1,t_2)$ for given $t_1$ and
$t_2$. But there exists a unitary time evolution operator
$U(-\infty,\infty) \equiv T \text {exp}\{i\int_{-\infty} ^{\infty}
d^4x {\cal L}_{int}(x)\}$. This can be easily understood since a
fermion, an anti-fermion and a gauge field can be
annihilated/created at different time. The possibility of state is
not conserved at a fixed time. But the integral of the possibility
over the time is conserved. In other words, the time evolution
operator $U(-\infty,\infty)$ is unitary.

Though the relativistic version of the solid quantization is Lorentz
invariant, the causality condition is different from the traditional
quantum field theory. For example for scalar field in local case,
the equal-time commutator $[\phi(\vec{x}),\pi(\vec{y})]$ equal zero
if $x$ and $y$ are spacelike separated, i.e.
$(x_0-y_0)^2-(\vec{x}-\vec{y})^2<0$. In non-local case, there exists
some possibility of non-zero commutator
$[\phi(\vec{x}),\pi(\vec{y})]$ for $-(\vec{x}-\vec{y})^2<0$. The
non-zero commutator is because of the space-time distribution of the
non-point fields. Therefore, the classic causality condition turns
into a quantum (possibility) condition. Approximately, one may think
the two non-point fields are spacelike separated if
$(x_0-y_0)^2-(\vec{x}-\vec{y})^2<-\vec{a}^2$, where $\vec{a}$ is the
size of the field.

In summary, we have proposed a new quantization - solid quantization
for non-point fields. The divergence in the loop integrals for point
particles needs to be taken care of with the regularization method.
This solid quantization condition is very natural and based on the
idea that a physical particle is not a mathematic point one. The
function in the commutation relations is another fundamental
properties of the particle as well as mass, spin, width, etc. The
divergence of loop integration could be systematically avoided from
the beginning. Both non-relativistic and relativistic version of
this solid quantization are given.

For the dimensional regularization, one has to use infinite
Lagrangian or bare Lagrangian to get finite physical results. This
method provide an interesting approach which is quite different from
traditional quantum field theory. In this paper, we did not specify
the function of $\Phi(\vec{x})$, $\Psi(\vec{p})$ or $H(p^2)$. To get
more information about the function of the particle, it is important
to do further numerical calculations to compare with the experiments
and traditional results.

\section*{Acknowledgements}

The author would like to thank Prof. Y. B. Dong for helpful
discussions.

\end{document}